\documentclass[final,authoryear,5p,times,twocolumn]{elsarticle}
\usepackage{epsfig}
\usepackage{amssymb}
\usepackage{float}
\journal{Journal of Theoretical Biology}

\begin{document}
\begin{frontmatter}

\title{Rewarding evolutionary fitness with links between populations promotes cooperation}

\author[hkt]{Zhen Wang}
\author[mfa]{Attila Szolnoki}
\author[umb]{Matja{\v z} Perc}
\address[hkt]{Department of Physics, Hong Kong Baptist University, Kowloon Tong, Hong Kong \& Center for Nonlinear Studies and Beijing-Hong Kong-Singapore Joint Center for Nonlinear and Complex systems, Institute of Computational and Theoretical Studies, Hong Kong Baptist University, Kowloon Tong, Hong Kong}
\address[mfa]{Institute of Technical Physics and Materials Science, Research Centre for Natural Sciences, Hungarian Academy of Sciences, H-1525 Budapest, Hungary}
\address[umb]{Faculty of Natural Sciences and Mathematics, University of Maribor, Koro{\v s}ka cesta 160, SI-2000 Maribor, Slovenia}

\begin{abstract}
Evolution of cooperation in the prisoner's dilemma and the public goods game is studied, where initially players belong to two independent structured populations. Simultaneously with the strategy evolution, players whose current utility exceeds a threshold are rewarded by an external link to a player belonging to the other population. Yet as soon as the utility drops below the threshold, the external link is terminated. The rewarding of current evolutionary fitness thus introduces a time-varying interdependence between the two populations. We show that, regardless of the details of the evolutionary game and the interaction structure, the self-organization of fitness and reward gives rise to distinguished players that act as strong catalysts of cooperative behavior. However, there also exist critical utility thresholds beyond which distinguished players are no longer able to percolate. The interdependence between the two populations then vanishes, and cooperators are forced to rely on traditional network reciprocity alone. We thus demonstrate that a simple strategy-independent form of rewarding may significantly expand the scope of cooperation on structured populations. The formation of links outside the immediate community seems particularly applicable in human societies, where an individual is typically member in many different social networks.
\end{abstract}


\end{frontmatter}

\section{Introduction}
Recent research has highlighted rewarding as an effective means to promote public cooperation
\citep{rand_s09, szolnoki_epl10, hauert_jtb10, mestertong_jtb11}. In comparison to peer \citep{fehr_n02, semmann_n03, de-quervain_s04, fowler_pnas05, hauert_s07, gaechter_s08, ohtsuki_n09, rockenbach_n09, deng_k_tpb12, vukov_pcbi13} and pool punishment \citep{sigmund_n10, szolnoki_pre11, perc_srep12, traulsen_prsb12}, rewarding may lead to higher total earnings without potential damage to reputation \citep{milinski_n02} or fear of retaliation \citep{dreber_n08}. The application of rewarding also avoids the problem of antisocial punishment \citep{herrmann_s08}, which has been shown to significantly challenge the effectiveness of sanctioning \citep{rand_jtb10, rand_nc11}. Although the majority of previous studies addressing the ``stick versus carrot'' dilemma \citep{sigmund_pnas01, hilbe_prsb10} concluded that punishment is more effective than rewarding in sustaining public cooperation \citep{sigmund_tee07}, evidence suggesting that rewards may be as effective as sanctions is mounting. Recent human experiments \citep{yamagishi_pnas12, egloff_pnas13} also challenge the strong reciprocity model \citep{fehr_hn02}, and related theoretical explorations \citep{szolnoki_prx13} indicate that the application of either reward or punishment, but not both, is evolutionary most advantageous.

Another relatively recent development is the study of evolutionary games on interdependent networks \citep{wang_z_epl12, gomez-gardenes_srep12, gomez-gardenes_pre12, wang_b_jsm12, wang_z_srep13b, jiang_ll_srep13, szolnoki_njp13}. The subject has gained on prominence after the discovery that even seemingly irrelevant changes in one network can have catastrophic and very much unexpected consequences in another network \citep{buldyrev_n10}. Since the evolution of cooperation, especially in human societies \citep{apicella_n12, rand_tcs13, helbing_n13}, also proceeds on such interdependent networks, it is therefore of interest to determine to what extent this interdependence influences the outcome of evolutionary games. It has been shown, for example, that biased utility functions suppress the feedback of individual success and lead to a spontaneous separation of time scales on interdependent networks \citep{wang_z_epl12}. If utilities are symmetric, cooperation is promoted by means of interdependent network reciprocity that relies on the simultaneous formation of correlated cooperative clusters on both networks \citep{wang_z_srep13}. In addition to these examples, non-trivial organization of cooperators across the interdependent layers \citep{gomez-gardenes_srep12}, strategy-independent information sharing \citep{szolnoki_njp13}, probabilistic interconnectedness \citep{wang_b_jsm12}, as well as optimal interdependence \citep{wang_z_srep13b}, have all been shown to extend the boundaries of traditional network reciprocity \citep{nowak_n92b} past its limits on isolated networks \citep{santos_pnas06, ohtsuki_n06, szabo_pr07, perc_jrsi13}.

Here we wish to extend the scope of evolutionary games on interdependent networks by introducing rewards for high-enough evolutionary fitness of individual players in the form of additional links that bridge the gap between two initially disconnected populations. We introduce a utility threshold $E$ that, if met or exceeded, allows the pertinent player to connect with the corresponding player in the other network. These rewards effectively introduce interdependence between the two populations, and they allow the rewarded players to increase their utility with a fraction of the utility of the player in the other population. However, as soon as the fitness of a player no longer reaches the threshold, its external link is terminated, although it may eventually be re-awarded if and when the utility of the player again becomes sufficiently large. Importantly, the on-off nature of the interdependence between the corresponding players in the two populations draws exclusively on the current level of fitness, without regard of previous evolutionary success or strategy. We consider the weak prisoner's dilemma game as representative for social dilemmas that are governed by pairwise interactions, and the public goods game which is representative for social dilemmas that are governed by group interactions. We also consider different types of networks to describe the interactions among players in each of the two structured populations. As we will show, regardless of these details, the self-organization of fitness and reward promotes the evolution of cooperation well past the boundaries imposed by traditional network reciprocity \citep{nowak_n92b}, as well as past the boundaries imposed by interdependent network reciprocity \citep{wang_z_srep13}, if only the utility threshold is sufficiently large. On the other hand, the threshold must not exceed a critical value, which could be well below the maximal possible utility a cooperator is able to reach if it would be fully surrounded by other cooperators. We will extend and explain these results in detail in Section 3, while in the subsequent section we proceed with the description of the studied evolutionary games.

\section{Evolutionary games}
The evolutionary games are staged on two disjoint square lattices or random regular graphs with periodic boundary conditions, each of size $N$, where initially each player $x$ is designated either as a cooperator $(s_x=C)$ or defector $(s_x=D)$ with equal probability. The weak prisoner's dilemma game is characterized by the temptation to defect $T = b$, reward for mutual cooperation $R = 1$, and both the punishment for mutual defection $P$ as well as the suckers payoff $S$ equaling $0$, where $1 < b \leq 2$ \citep{nowak_n92b}. In this case a player receives its payoff by playing the game with all its neighbors. For the public goods game, players are arranged into overlapping groups of size $G$, where every player is thus surrounded by its $k=G-1$ neighbors and is a member in $g=G$ different groups \citep{santos_n08, perc_jrsi13}. In each group, cooperators contribute $1$ to the public good, while defectors contribute nothing. The sum of contributions is subsequently multiplied by the factor $r>1$, reflecting the synergetic effects of cooperation, and the resulting amount is equally shared amongst the $G$ group members. Here the total payoff of a player is the sum of payoffs from all the $g$ groups where she is member.

We simulate the evolutionary process on both networks in accordance with the standard Monte Carlo simulation procedure comprising the following elementary steps. First, a player $x$ is selected randomly and its payoff $\Pi_x$ is determined based on the governing evolutionary game (either the weak prisoner's dilemma game or the public goods game). Next, a neighbor $y$ from the same network is chosen randomly and acquires its payoff $\Pi_y$ in the same way. Lastly, player $y$ adopts the strategy of player $x$ with the probability
\begin{equation}
W(s_{x} \rightarrow s_{y})=\frac{1}{1+\exp[(U_y-U_x)/K]} \,,
\label{fermi}
\end{equation}
where $K=0.1$ quantifies the uncertainty related to the strategy adoption process \citep{szabo_pr07}, while $U_x$ and $U_y$ are the utilities of players $x$ and $y$, respectively. All those players that have an external link to the corresponding player $x^{\prime}$ in the other network have the utility
\begin{equation}
U_x = \Pi_x + \alpha \Pi_{x^{\prime}}\,,
\label{utility}
\end{equation}
while those that do not have an external link retain $U_x=\Pi_x$. We emphasize at this point that the external links are directed. Hence, only player $x$ benefits from the additional link, but not player $x^{\prime}$. We also do not allow a direct interaction between the two mentioned players. Based on our preceding work \citep{wang_z_srep13b}, where we have studied the general impact of the value of $\alpha$ and the related optimal interdependence between two networks, we here use a fixed value of $\alpha=0.5$ without loosing generality. Monte Carlo simulations are performed on sufficiently large networks ranging in size from $N=4\cdot 10^4$ to $2.5\cdot 10^5$ near transition points to avoid accidental extinction of the two competing strategies. The stationary fraction of cooperators $\rho_C$ is recorded after the system reaches dynamical equilibrium, i.e., when the average cooperation level becomes time independent. More specifically, we perform $10^4$ Monte Carlo steps (MCS) to reach the stationary state, and subsequently $10^6$ more steps to record $\rho_C$. Moreover, we average the final outcome over up to $100$ independent initial conditions to further improve accuracy.

To distinguish who is eligible for an external link, we introduce a two-value tag for each player. If $q_x=1$ the utility of the player $x$ is determined according to Eq.~\ref{utility}, while if $q_x=0$ the utility remains equal to the payoff stemming from the interactions on the host network. Initially all players have $q_x=0$. Subsequently, if the utility of player $x$ with its current tag value is equal or above the threshold $E$, we set $q_x=1$. If not, the tag remains zero. Evidently, if the value of $E$ is minimal, which is $E=0$ for the weak prisoner's dilemma game, then all players are rewarded with an external link and hence are able to enhance their utility according to Eq.~\ref{utility}. If the value of $E$ is too high, on the other hand, no player is ever awarded with an external link, and the two populations remain disconnected (independent of one another). Intermediate values of $E$, however, promise interesting results as the evolutionary processes on each individual network become interdependent through the time-varying addition and removal of the external links. We note that the selection of an external player from the other population is done randomly and is not influenced by its actual tag. If the linking would be tag-based, this might give rise to further nontrivial effects due to the establishment of bilateral connections, as it discussed recently in a related work by Fu et al. \citep{fu_srep12}. While we wish to avoid this possibility within the scope of present study, it may constitute a worthy research challenge to be addressed in the future.

\section{Results}

\begin{figure}
\centerline{\epsfig{file=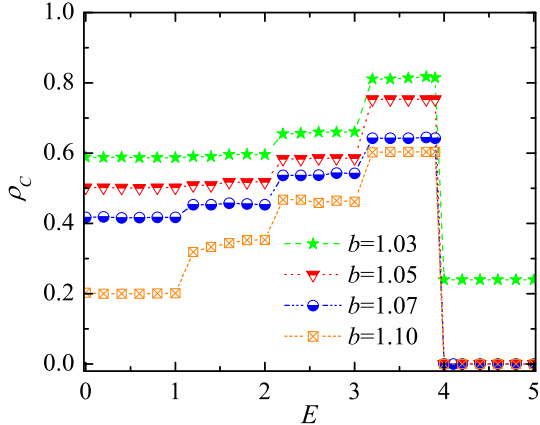,width=7cm}}
\caption{\label{pd_sq} Intermediate values of the utility threshold $E$ are able to sustain widespread cooperation in the evolutionary prisoner's dilemma game, even at temptations to defect where the evolution on an isolated graph would terminate in an absorbing $D$ phase (see legend). However, as soon as a critical value $E=E_c$ is reached, the two populations become independent, and the value of $\rho_C$ drops to the level that is characteristic for an isolated network. Since here the interaction network for both populations is a square lattice where every player is connected to its four nearest neighbors, the critical value is $E_c=4$, which is the maximal payoff of a cooperator (if all its neighbors are also cooperators). We note, however, that this is accurate threshold is an exception, since in general the value of $E_c$ is somewhat lower than the maximal cooperator payoff, as it also depends on the typical size of cooperative clusters [which in turn depends on $b$ ($r$ in the public goods game)]. At $E=0$, on the other hand, the two populations are completely interdependent (every player has a link to the corresponding player in the other network), and due to interdependent network reciprocity \citep{wang_z_srep13}, the level of cooperation is still significantly higher than on an isolated network (as obtained for $E \geq E_c$).}
\end{figure}

For easier comparison, we first present results for the most traditional setup, using as the interaction network for both populations the square lattice and the weak prisoner's dilemma as the governing evolutionary game. Results presented in Fig.~\ref{pd_sq} illustrate the threshold-dependence of the cooperation level, as recorded in both populations together. It can be observed that there exist an intermediate value of the utility threshold $E$, at which the evolution of cooperation is optimally promoted. However, while the rise of $\rho_C$ is steady as $E$ increases, the drop is sudden, and it occurs at a critical values of the threshold ($E=E_c=4$), which is related to the maximal payoff attainable by a cooperator. In the depicted case the critical value is actually identical to the maximal cooperator payoff, but as we will show in what follows, this is an exception rather than the rule. The $E=0$ threshold renders all players worthy of the reward [an external link to the corresponding player in the other network and with it related higher utility (see Eq.~\ref{utility})], and thus introduces full interdependence between the two populations. The evolution of cooperation then proceeds with the support of interdependent network reciprocity \citep{wang_z_srep13}. On the other hand, if $E \geq E_c$, players are in general unable to reach the threshold, which leaves the two populations fully independent. Traditional network reciprocity \citep{nowak_n92b} is then the sole mechanism supporting the survival of cooperators, and it can be observed that it is less effective than interdependent network reciprocity (the $E=0$ case). Just slightly below the critical value, however, cooperators fare best, and in what follows we will present results in favor of the robustness of this observation, as well as results that explain it.

\begin{figure}
\centerline{\epsfig{file=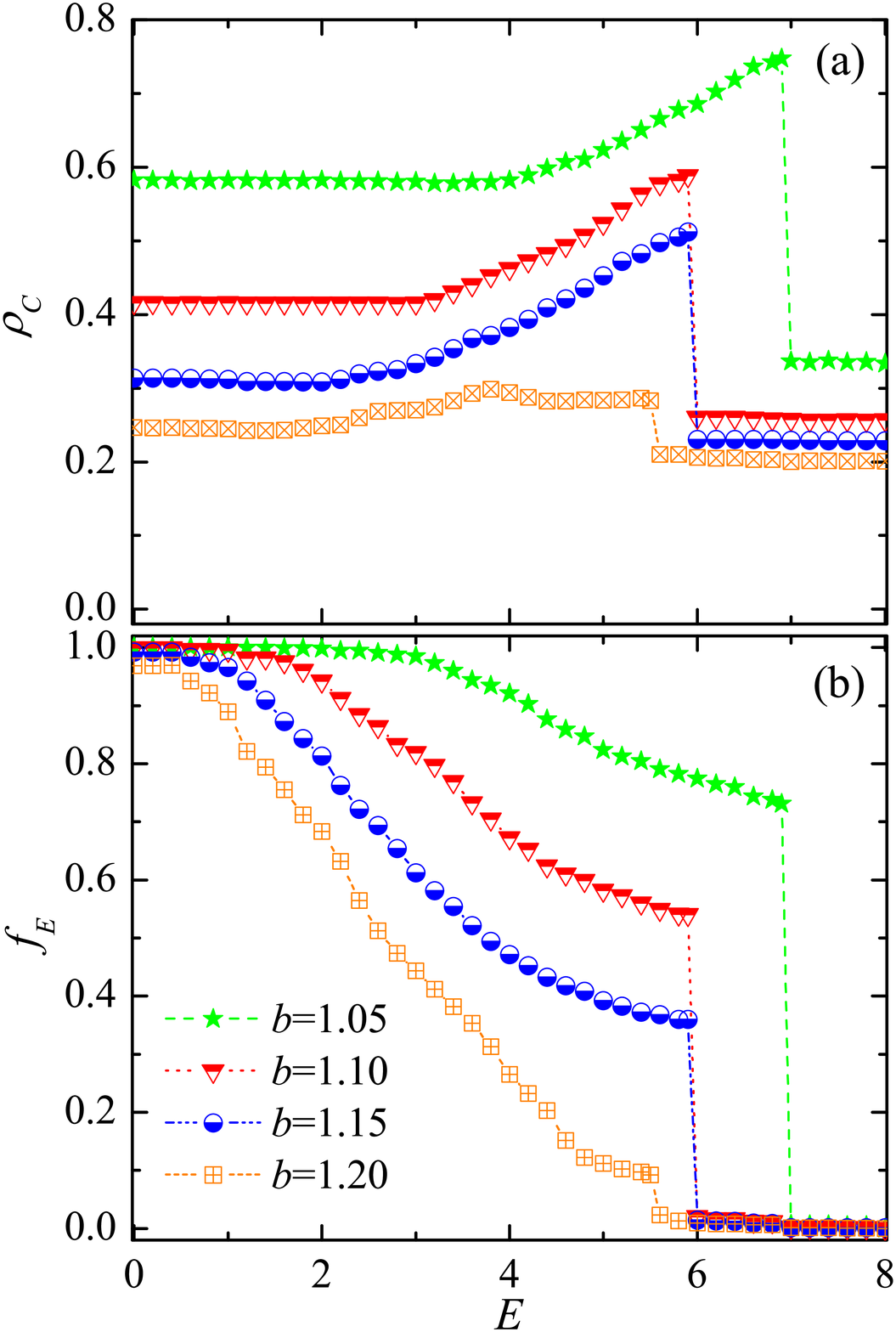,width=7cm}}
\caption{\label{random} Promotion of cooperation at intermediate values of the utility threshold $E$ is robust against the variation of the interaction network. The upper layer depicts the fraction of cooperators $\rho_C$ in dependence on $E$ for different values of $b$ (see legend), as obtained on a regular random graph where every player has degree $k=8$. The bottom layer depicts the fraction of rewarded players $f_E$ whose utility is above the threshold $E$. Note that there is a strong correlation between the dependence of $\rho_C$ and $f_E$. Importantly, here the critical value of $E=E_c$ depends somewhat on $b$ (because the values of $b$ determines the typical size of cooperative clusters), and it is also below the maximally attainable payoff of a cooperator (if all its neighbors would also be cooperators). This indicates that the $E_c$ value does not necessarily imply a strict prohibition of external links. Cooperators that are located deep within large cooperative clusters may still be eligible for the reward, but the cooperators at the edge are vulnerable, which will eventually lead to the erosion of cooperative clusters and the disappearance of external links even at $E_c$ values that are smaller than the maximal cooperator payoff. And since larger values of $b$ give rise to smaller cooperative clusters, the value of $E_c$ decreases as the temptation to defect increases.}
\end{figure}

We first test the robustness by replacing the square lattice interaction topology with the random regular graph where every player has eight neighbors. Results presented in Fig.~\ref{random}(a) are qualitatively identical to those presented in Fig.~\ref{pd_sq}. The dependence of $\rho_C$ on $E$ is somewhat smoother, and due to the larger number of neighbors (on the square lattice players have only four neighbors each), the critical threshold value $E_c$ could be approximately two times as large as on the square lattice. But not quite. Indeed, it is lower than the maximally attainable cooperator payoff (which in this case would be eight), and it also decreases as the temptation to defect $b$ increases. These subtle features, which remain hidden in Fig.~\ref{pd_sq}, provide vital insights that help to understand why and how the evolution of cooperation is promoted. In particular, the fact that the value of $E_c$ is lower than the maximal payoff of a cooperator indicates that the critical threshold value does not necessarily involves the prohibition of rewarding. Cooperators deep in the bulk of a cooperative domain may qualify at any given time, but those along the interface are vulnerable because they are unable to reach a sufficiently high fitness to be rewarded. Defectors are therefore able to invade, which eventually leads to the disintegration of the clusters. In the stationary state even the aforementioned ``central'' cooperators will fall victim to the high utility threshold, and they will not be awarded an external link, and thus also will not be able to resist the invasion of defectors. To support this argument and to demonstrate the key role of the interdependence between the two populations, we present in Fig.~\ref{random}(b) the fraction of players that have an external link (are rewarded) in dependence on the threshold $E$. It can be observed that there is a direct correlation between the fraction of rewarded players and the fraction of cooperators in the population.

\begin{figure}
\centerline{\epsfig{file=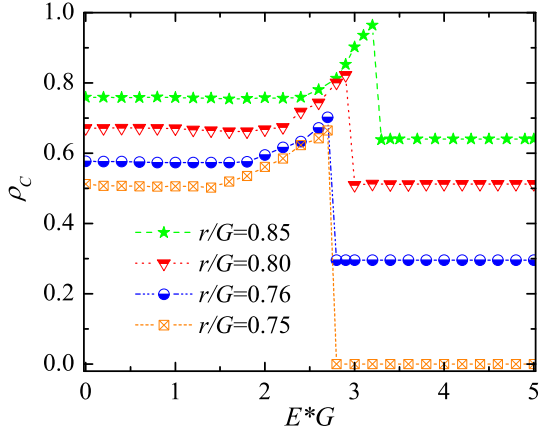,width=7cm}}
\caption{\label{pgg} Promotion of cooperation at intermediate values of the utility threshold $E$ is robust against the variation of game details. Here results for the public goods game are presented, as obtained on a square lattice where every player is involved in five groups of size $G=5$. As in Fig.~\ref{random}, the critical value of the threshold $E=E_c$ depends on the multiplication factor $r$ (see legend), and it is lower than the maximally attainable payoff of a cooperator in five groups where all other players would also cooperate. As in the prisoner's dilemma game, this is related to the decreasing size of cooperative clusters as $r$ decreases, and to the vulnerability of cooperators that are positioned along the edges of cooperative domains (see also Fig.~\ref{random}).}
\end{figure}

Going further in the exploration of robustness of our findings, we return to the square lattice as the interaction network, but we replace the pairwise driven prisoner's dilemma game with the group-driven public goods game \citep{santos_n08, perc_jrsi13}. As Fig.~\ref{pgg} illustrates, in this case too the evolution of public cooperation is promoted, and essentially the same conclusions apply as outlined above for the results presented in Fig.~\ref{random}. In agreement with the behavior we have observed for the prisoner's dilemma game, the harsher the social dilemma (in this case the smaller the multiplication factor $r$), the smaller the critical threshold value $E=E_c$ at which the two populations become fully independent and the cooperation level drops. Just like larger values of $b$ in the prisoner's dilemma game decrease the typical size of compact cooperative clusters, so do decreasing values of $r$ in the public goods game do the same. Both the increase of the temptation to defect and the decrease of the multiplication factor enhance the severity of the social dilemma. And because the cooperative clusters become smaller in size, their ability to contain cooperators that might be able reach close-to-optimal payoffs decreases, and so does the value of $E_c$.

\begin{figure}
\centerline{\epsfig{file=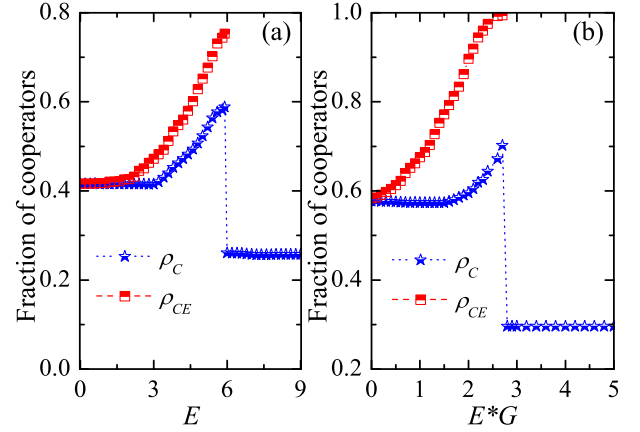,width=7.96cm}}
\caption{\label{fc} The evolution of cooperation among players with an external link explains the promotion of cooperation at the microscopic level. Depicted is the fraction of cooperators amongst all the players that fulfil $U_x \geq E$, denoted as $\rho_{CE}$, and the overall fraction of cooperators in both populations, in dependence on the threshold $E$. Only if the distinguished players are rare (which occurs at sufficiently high values of $E$), but at the same time are also still able to percolate (which requires $E<E_c$), will cooperative behavior be optimally promoted. Left panel shows the results obtained with the prisoner's dilemma game on a regular random graph where every player has degree $k=8$, and using the temptation to defect $b=1.1$. Right panel depicts the results of the public goods game on a square lattice using $r/G=0.76$. We note that to plot $\rho_{CE}$ for $E \geq E_c$ is meaningless, because no such players exist, as evidenced by the results presented in Fig.~\ref{random}(b) for the prisoner's dilemma game.}
\end{figure}

To explain the observed behavior at the microscopic level, we measure the cooperation level specifically amongst those players who possess an external link because their utility is above the threshold value $E$. In Fig.~\ref{fc}, we plot this quantity (denoted by $\rho_{CE}$), and we also show the cooperation level in both populations for comparison. If the value of $E$ is small, then there is no significant difference between $\rho_{CE}$ and $\rho_E$. If we increase the threshold, then two mechanisms will emerge that support each other's impact. On the one hand, by increasing the value of $E$, heterogeneity is introduced to both populations because not all players will fulfill the condition to be rewarded with an external link. Such heterogeneity will enhance the impact of network reciprocity because of the long-term advantage of cooperative leaders (the latter was identified earliest with the discovery that scale-free networks provide a unifying framework for the evolution of cooperation \citep{santos_prl05}). The distinguished cooperators will be followed by those who fail to reach the rewarding threshold, which will in turn increase $\rho_C$. On the other hand, the relatively rare rewarded players need to percolate, i.e., they still have to be sufficiently common for their influence in the population to overlap. If their influence does not overlap, in the absence of percolation, the advantage of cooperative behavior cannot manifest, which will reduce the cooperation level. Noteworthy, this behavior is related to the optimal density of players on a network, which we have recently explored in isolated structured populations \citep{wang_z_pre12b, wang_z_srep12}.

\begin{figure}
\centerline{\epsfig{file=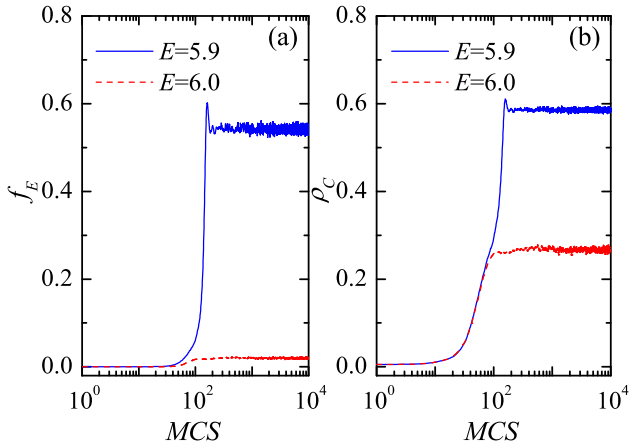,width=8.2cm}}
\caption{\label{direct} Inserting a small cooperative domain, consisting of $780$ players, into a homogeneous populations of $4\cdot 10^4$ defectors reveals the spreading of distinguished (rewarded) players, which is due to the self-organization of fitness and reward. Depicted is the time evolution of the fraction of rewarded players $f_E$ that satisfy $U_x \geq E$ (left) and the overall cooperator density $\rho_C$ (right) versus the number of Monte Carlo steps for $b=1.1$, $E=5.9$ (solid blue line) and $E=6.0$ (dashed red line), as obtained on a regular random graph where every player has degree $k=8$. If the threshold is higher than the critical value $E_c$ ($E=6.0$ in this case), a negligible minority of players is rewarded, but they are unable to percolate. Accordingly, cooperators are forced to rely solely on traditional network reciprocity, and their density is accordingly low. Just a tiny drop in $E$ below $E_c$ ($E=5.9$ in this case) enables rewarded players to percolate and thus to support each other, and therefore the stationary fraction of cooperators rises significantly.}
\end{figure}

Finally, to provide direct evidence that the self-organized spreading of rewarded players will elevate cooperation, we devise a specific experiment, wherein, starting from a full $D$ state, we insert a small cooperator domain into both graphs which are not correlated to each other. To illustrate the robustness of the leading mechanism we use random regular graphs, where the cooperative domain is constructed by starting from a single cooperative player and subsequently denoting all its neighbors and the neighbors of neighbors, and so forth, as cooperators, until the initial fraction of cooperators in the whole population reaches $2\%$. By using two different threshold values, we monitor how the fraction of rewarded players, and how the general cooperation level, evolve. To emphasize the delicate role of the threshold, we have chosen two $E$ values that are slightly below and above the critical utility $E_c$ for the applied $b$ value. As Fig.~\ref{direct} illustrates, players cannot become distinguished if the demand for the reward is too high, and consequently the cooperation can reach a level which is attainable solely by traditional network reciprocity. If we lower the threshold only slightly, then more players become eligible for the reward and can build an external link to the corresponding player in the other network, and the latter will in turn support the higher utility in future trails. If we compare the applied values of $E$ in Fig.~\ref{direct}, it becomes clear than just a tiny change can yield a significant difference in $f_E$ (the fraction of players that are rewarded). Because the rare, yet still sufficiently common, rewarded players percolate, they can support each other through the overlap of extensive cooperative clusters that surround them, which in turn significantly elevates the cooperation level in the whole system.

\section{Discussion}
The evolution of cooperation among unrelated individuals is a long-standing problem which attracts research from a wide range fields, including biology and the social sciences \citep{mestertong_qrb92, mestertong_01, nowak_s06, sigmund_10}. We have studied the evolution of cooperation in two initially independent structured populations, where players were rewarded with an external link to a player in the other network if their current utility reached or exceeded a given threshold. Being rewarded with an external link implied the acquisition of part of the corresponding player's payoff, and thus increasing one's own evolutionary potential in the next round of the game. Importantly, we have considered rewards being representative only for the current evolutionary success. Thus, as soon as the utility of a player dropped below the utility threshold, the external link was removed.

It was shown that this simple but intuitive coevolutionary rule introduces a time-varying interdependence between the two populations, which provides a dynamical coupling between them. As a result of coevolution between strategy dependent fitness and external links a threshold-dependent interdependency will emerge and stabilize. We have also shown that it may significantly promote the evolution of cooperation if the applied utility threshold is neither too small nor to large. If it is too small, the two populations are fully interdependent, and the benefits for the evolution of cooperation are similar to those warranted by interdependent network reciprocity \citep{wang_z_srep13}. If, on the other hand, the utility threshold is too high, the two populations are fully independent, and cooperators may take advantage only of traditional network reciprocity \citep{nowak_n92b}. For intermediate values of the utility threshold, and especially for values close but lower than a critical threshold, cooperators can take advantage of percolating distinguished players that act as strong catalysts of cooperative behavior by offering a cooperation supporting mechanism to evolve at a higher level. Fine-tuning the utility threshold has the important role of adjusting the density of rewarded players just right. If the rewarded players are too many, the heterogeneity in the population vanishes, and so does the additional support for the evolution of cooperation. If the rewarded players are too few, they are unable to percolate. Indeed, the rewarded players must be rare, but at the same time frequent enough to percolate for the optimal conditions to emerge. The percolation is related to the optimal density of players on a network, which has recently been explored in detail on isolated populations for pairwise social dilemmas and the public goods game \citep{wang_z_pre12b}.

We have also established that the value of the critical threshold is related to the maximal payoff that is attainable by a cooperator, although only in special cases will it be exactly equal to this value (see Fig.~\ref{pd_sq}). In general, the critical utility threshold is lower that the maximal cooperator payoff because cooperators that are located at the interfaces of cooperative domains can never reach the highest payoffs. At such high threshold values (equal to the maximal cooperator payoff), the interfaces are therefore vulnerable to the invasion by defectors, which eventually leads to the dissolution of cooperative clusters. The value of the critical utility threshold also decreases with the increasing severity of the social dilemma (increasing value of $b$ in the prisoner's dilemma game, or the decreasing value of the multiplication factor $r$ in the public goods game), which is due to the fact that at harsher conditions the cooperative clusters become smaller. Their potential to harbor cooperators that might reach a close-to-optimal payoff thus decrease as well, and accordingly so does the critical utility threshold that still warrants percolation of those that are rewarded with an external link.

All the presented results are independent of the structure of the applied interaction networks and the studied social dilemma games, and thus appear to have a high degree of universality. We hope that the demonstration of the spontaneous emergence of optimal interdependence by means of a simple coevolutionary rule based on the self-organization of reward and fitness will attract further research on the evolution of cooperation on interdependent networks \citep{wang_z_epl12, gomez-gardenes_srep12, gomez-gardenes_pre12, wang_b_jsm12, wang_z_srep13, wang_z_srep13b, jiang_ll_srep13, szolnoki_njp13}. The consideration of interdependence appears to be particularly relevant for human societies \citep{apicella_n12, rand_tcs13, helbing_n13}, where individuals are typically members in many different networks, and those networks might play different roles in the evolutionary process. Multilayer networks are already in the focus of attention as the most apt description of a networked society \citep{kivela_ax13}, and we hope that our study will help contribute to the continued vibrancy of this research avenue.

\section*{Acknowledgments}
This research was supported by the Hungarian National Research Fund (Grant K-101490), TAMOP-4.2.2.A-11/1/KONV-2012-0051, and the Slovenian Research Agency (Grant J1-4055).


\begin{thebibliography}{57}
\expandafter\ifx\csname natexlab\endcsname\relax\def\natexlab#1{#1}\fi
\expandafter\ifx\csname url\endcsname\relax
  \def\url#1{\texttt{#1}}\fi
\expandafter\ifx\csname urlprefix\endcsname\relax\def\urlprefix{URL }\fi

\bibitem[{Apicella et~al.(2012)Apicella, Marlowe, Fowler, and
  Christakis}]{apicella_n12}
Apicella, C.~L., Marlowe, F.~W., Fowler, J.~H., Christakis, N.~A., 2012. Social
  networks and cooperation in hunter-gatherers. Nature 481, 497--501.

\bibitem[{Buldyrev et~al.(2010)Buldyrev, Parshani, Paul, Stanley, and
  Havlin}]{buldyrev_n10}
Buldyrev, S.~V., Parshani, R., Paul, G., Stanley, H.~E., Havlin, S., 2010.
  Catastrophic cascade of failures in interdependent networks. Nature 464,
  1025--1028.

\bibitem[{de~Quervain et~al.(2004)de~Quervain, Fischbacher, Treyer,
  Schellhammer, Schnyder, Buck, and Fehr}]{de-quervain_s04}
de~Quervain, D. J.-F., Fischbacher, U., Treyer, V., Schellhammer, M., Schnyder,
  U., Buck, A., Fehr, E., 2004. The neural basis of altruistic punishments.
  Science 305, 1254--1258.

\bibitem[{Deng et~al.(2012)Deng, Li, Kurokawa, and Chu}]{deng_k_tpb12}
Deng, K., Li, Z., Kurokawa, S., Chu, T., 2012. Rare but severe concerted
  punishment that favors cooperation. Theor. Popul. Biol. 81, 284--291.

\bibitem[{Dreber et~al.(2008)Dreber, Rand, Fudenberg, and Nowak}]{dreber_n08}
Dreber, A., Rand, D.~G., Fudenberg, D., Nowak, M.~A., 2008. Winners don't
  punish. Nature 452, 348--351.

\bibitem[{Egloff et~al.(2013)Egloff, Richter, and Schmukle}]{egloff_pnas13}
Egloff, B., Richter, D., Schmukle, S., 2013. Need for conclusive evidence that
  positive and negative reciprocity are unrelated. Proc. Natl. Acad. Sci. USA
  110, E786.

\bibitem[{Fehr et~al.(2002)Fehr, Fischbacher, and G{\"a}chter}]{fehr_hn02}
Fehr, E., Fischbacher, U., G{\"a}chter, S., 2002. Strong reciprocity, human
  cooperation and the enforcement of social norms. Human Nature 13, 1--25.

\bibitem[{Fehr and G{\"a}chter(2002)}]{fehr_n02}
Fehr, E., G{\"a}chter, S., 2002. Altruistic punishment in humans. Nature 415,
  137--140.

\bibitem[{Fowler(2005)}]{fowler_pnas05}
Fowler, J.~H., 2005. Altruistic punishment and the origin of cooperation. Proc.
  Natl. Acad. Sci. USA 102, 7047--7049.

\bibitem[{Fu et~al.(2012)Fu, Tarnita, Christakis, Wang, Rand, and
  Nowak}]{fu_srep12}
Fu, F., Tarnita, C., Christakis, N., Wang, L., Rand, D., Nowak, M., 2012.
  Evolution of in-group favoritism. Sci. Rep. 2, 460.

\bibitem[{G{\"a}chter et~al.(2008)G{\"a}chter, Renner, and
  Sefton}]{gaechter_s08}
G{\"a}chter, S., Renner, E., Sefton, M., 2008. The long-run benefits of
  punishment. Science 322, 1510.

\bibitem[{G{\'o}mez-Garde{\~n}es
  et~al.(2012{\natexlab{a}})G{\'o}mez-Garde{\~n}es, Gracia-L{\'a}zaro, Flor{\'
  \i}a, and Moreno}]{gomez-gardenes_pre12}
G{\'o}mez-Garde{\~n}es, J., Gracia-L{\'a}zaro, C., Flor{\' \i}a, L.~M., Moreno,
  Y., 2012{\natexlab{a}}. Evolutionary dynamics on interdependent populations.
  Phys. Rev. E 86, 056113.

\bibitem[{G{\'o}mez-Garde{\~n}es
  et~al.(2012{\natexlab{b}})G{\'o}mez-Garde{\~n}es, Reinares, Arenas, and
  Flor{\' \i}a}]{gomez-gardenes_srep12}
G{\'o}mez-Garde{\~n}es, J., Reinares, I., Arenas, A., Flor{\' \i}a, L.~M.,
  2012{\natexlab{b}}. Evolution of cooperation in multiplex networks. Sci. Rep.
  2, 620.

\bibitem[{Hauert(2010)}]{hauert_jtb10}
Hauert, C., 2010. Replicator dynamics of reward \& reputation in public goods
  games. J. Theor. Biol. 267, 22--28.

\bibitem[{Hauert et~al.(2007)Hauert, Traulsen, Brandt, Nowak, and
  Sigmund}]{hauert_s07}
Hauert, C., Traulsen, A., Brandt, H., Nowak, M.~A., Sigmund, K., 2007. Via
  freedom to coercion: The emergence of costly punishment. Science 316,
  1905--1907.

\bibitem[{Helbing(2013)}]{helbing_n13}
Helbing, D., 2013. Globally networked risks and how to respond. Nature 497,
  51--59.

\bibitem[{Herrmann et~al.(2008)Herrmann, Thoni, and G{\"a}chter}]{herrmann_s08}
Herrmann, B., Thoni, C., G{\"a}chter, S., 2008. Antisocial punishment across
  societies. Science 319, 1362--1367.

\bibitem[{Hilbe and Sigmund(2010)}]{hilbe_prsb10}
Hilbe, C., Sigmund, K., 2010. Incentives and opportunism: from the carrot to
  the stick. Proc. R. Soc. B 277, 2427--2433.

\bibitem[{Jiang and Perc(2013)}]{jiang_ll_srep13}
Jiang, L.-L., Perc, M., 2013. Spreading of cooperative behaviour across
  interdependent groups. Sci. Rep. 3, 2483.

\bibitem[{Kivel{\"a} et~al.(2013)Kivel{\"a}, Arenas, Barthelemy, Gleeson,
  Moreno, and Porter}]{kivela_ax13}
Kivel{\"a}, M., Arenas, A., Barthelemy, M., Gleeson, J.~P., Moreno, Y., Porter,
  M.~A., 2013. Multilayer networks. arXiv:1309.7233.

\bibitem[{Mesterton-Gibbons(2001)}]{mestertong_01}
Mesterton-Gibbons, M., 2001. An Introduction to Game-Theoretic Modelling, 2nd
  Edition. American Mathematical Society, Providence, RI.

\bibitem[{Mesterton-Gibbons and Dugatkin(1992)}]{mestertong_qrb92}
Mesterton-Gibbons, M., Dugatkin, L.~A., 1992. Cooperation among unrelated
  individuals: Evolutionary factors. The Quarterly Review of Biology 67,
  267--281.

\bibitem[{Mesterton-Gibbons et~al.(2011)Mesterton-Gibbons, Gavrilets, Gravner,
  and Akcay}]{mestertong_jtb11}
Mesterton-Gibbons, M., Gavrilets, S., Gravner, J., Akcay, E., 2011. Models of
  coalition or alliance formation. J. Theor. Biol. 274, 187--204.

\bibitem[{Milinski et~al.(2002)Milinski, Semmann, and Krambeck}]{milinski_n02}
Milinski, M., Semmann, D., Krambeck, H.-J., 2002. Reputation helps to solve the
  'tragedy of the commons'. Nature 415, 424--426.

\bibitem[{Nowak(2006)}]{nowak_s06}
Nowak, M.~A., 2006. Five rules for the evolution of cooperation. Science 314,
  1560--1563.

\bibitem[{Nowak and May(1992)}]{nowak_n92b}
Nowak, M.~A., May, R.~M., 1992. Evolutionary games and spatial chaos. Nature
  359, 826--829.

\bibitem[{Ohtsuki et~al.(2006)Ohtsuki, Hauert, Lieberman, and
  Nowak}]{ohtsuki_n06}
Ohtsuki, H., Hauert, C., Lieberman, E., Nowak, M.~A., 2006. A simple rule for
  the evolution of cooperation on graphs and social networks. Nature 441,
  502--505.

\bibitem[{Ohtsuki et~al.(2009)Ohtsuki, Iwasa, and Nowak}]{ohtsuki_n09}
Ohtsuki, H., Iwasa, Y., Nowak, M.~A., 2009. Indirect reciprocity provides only
  a narrow margin of efficiency for costly punishment. Nature 457, 79--82.

\bibitem[{Perc(2012)}]{perc_srep12}
Perc, M., 2012. Sustainable institutionalized punishment requires elimination
  of second-order free-riders. Sci. Rep. 2, 344.

\bibitem[{Perc et~al.(2013)Perc, G{\'o}mez-Garde{\~n}es, Szolnoki, and
  Flor{\'{\i}a and Y. Moreno}}]{perc_jrsi13}
Perc, M., G{\'o}mez-Garde{\~n}es, J., Szolnoki, A., Flor{\'{\i}a and Y.
  Moreno}, L.~M., 2013. Evolutionary dynamics of group interactions on
  structured populations: a review. J. R. Soc. Interface 10, 20120997.

\bibitem[{Rand and Nowak(2013)}]{rand_tcs13}
Rand, D.~A., Nowak, M.~A., 2013. Human cooperation. Trends in Cognitive
  Sciences 17, 413--425.

\bibitem[{Rand et~al.(2010)Rand, Armao, Nakamaru, and Ohtsuki}]{rand_jtb10}
Rand, D.~G., Armao, J.~J., Nakamaru, M., Ohtsuki, H., 2010. Anti-social
  punishment can prevent the co-evolution of punishment and cooperation. J.
  Theor. Biol. 265, 624--632.

\bibitem[{Rand et~al.(2009)Rand, Dreber, Ellingsen, Fudenberg, and
  Nowak}]{rand_s09}
Rand, D.~G., Dreber, A., Ellingsen, T., Fudenberg, D., Nowak, M.~A., 2009.
  Positive interactions promote public cooperation. Science 325, 1272--1275.

\bibitem[{Rand and Nowak(2011)}]{rand_nc11}
Rand, D.~G., Nowak, M.~A., 2011. The evolution of antisocial punishment in
  optional public goods games. Nat. Commun. 2, 434.

\bibitem[{Rockenbach and Milinski(2009)}]{rockenbach_n09}
Rockenbach, B., Milinski, M., 2009. How to treat those of ill repute. Nature
  457, 39--40.

\bibitem[{Santos and Pacheco(2005)}]{santos_prl05}
Santos, F.~C., Pacheco, J.~M., 2005. Scale-free networks provide a unifying
  framework for the emergence of cooperation. Phys. Rev. Lett. 95, 098104.

\bibitem[{Santos et~al.(2006)Santos, Pacheco, and Lenaerts}]{santos_pnas06}
Santos, F.~C., Pacheco, J.~M., Lenaerts, T., 2006. Evolutionary dynamics of
  social dilemmas in structured heterogeneous populations. Proc. Natl. Acad.
  Sci. USA 103, 3490--3494.

\bibitem[{Santos et~al.(2008)Santos, Santos, and Pacheco}]{santos_n08}
Santos, F.~C., Santos, M.~D., Pacheco, J.~M., 2008. Social diversity promotes
  the emergence of cooperation in public goods games. Nature 454, 213--216.

\bibitem[{Semmann et~al.(2003)Semmann, Krambeck, and Milinski}]{semmann_n03}
Semmann, D., Krambeck, H.-J., Milinski, M., 2003. Volunteering leads to
  rock-paper-scissors dynamics in a public goods game. Nature 425, 390--393.

\bibitem[{Sigmund(2007)}]{sigmund_tee07}
Sigmund, K., 2007. Punish or perish? retailation and collaboration among
  humans. Trends Ecol. Evol. 22, 593--600.

\bibitem[{Sigmund(2010)}]{sigmund_10}
Sigmund, K., 2010. The Calculus of Selfishness. Princeton University Press,
  Princeton, NJ.

\bibitem[{Sigmund et~al.(2010)Sigmund, De~Silva, Traulsen, and
  Hauert}]{sigmund_n10}
Sigmund, K., De~Silva, H., Traulsen, A., Hauert, C., 2010. Social learning
  promotes institutions for governing the commons. Nature 466, 861--863.

\bibitem[{Sigmund et~al.(2001)Sigmund, Hauert, and Nowak}]{sigmund_pnas01}
Sigmund, K., Hauert, C., Nowak, M.~A., 2001. Reward and punishment. Proc. Natl.
  Acad. Sci. USA 98, 10757--10762.

\bibitem[{Szab{\'o} and F{\'a}th(2007)}]{szabo_pr07}
Szab{\'o}, G., F{\'a}th, G., 2007. Evolutionary games on graphs. Phys. Rep.
  446, 97--216.

\bibitem[{Szolnoki and Perc(2010)}]{szolnoki_epl10}
Szolnoki, A., Perc, M., 2010. Reward and cooperation in the spatial public
  goods game. EPL 92, 38003.

\bibitem[{Szolnoki and Perc(2013{\natexlab{a}})}]{szolnoki_prx13}
Szolnoki, A., Perc, M., 2013{\natexlab{a}}. Correlation of positive and
  negative reciprocity fails to confer an evolutionary advantage: Phase
  transitions to elementary strategies. Phys. Rev. X 3, 041021.

\bibitem[{Szolnoki and Perc(2013{\natexlab{b}})}]{szolnoki_njp13}
Szolnoki, A., Perc, M., 2013{\natexlab{b}}. Information sharing promotes
  prosocial behaviour. New J. Phys. 15, 053010.

\bibitem[{Szolnoki et~al.(2011)Szolnoki, Szab{\'o}, and Perc}]{szolnoki_pre11}
Szolnoki, A., Szab{\'o}, G., Perc, M., 2011. Phase diagrams for the spatial
  public goods game with pool punishment. Phys. Rev. E 83, 036101.

\bibitem[{Traulsen et~al.(2012)Traulsen, R{\"o}hl, and
  Milinski}]{traulsen_prsb12}
Traulsen, A., R{\"o}hl, T., Milinski, M., 2012. An economic experiment reveals
  that humans prefer pool punishment to maintain the commons. Proc. R. Soc. B
  279, 3716--3721.

\bibitem[{Vukov et~al.(2013)Vukov, Pinheiro, Santos, and
  Pacheco}]{vukov_pcbi13}
Vukov, J., Pinheiro, F., Santos, F., Pacheco, J., 2013. Reward from punishment
  does not emerge at all costs. PLoS Comput. Biol. 9, e1002868.

\bibitem[{Wang et~al.(2012{\natexlab{a}})Wang, Chen, and Wang}]{wang_b_jsm12}
Wang, B., Chen, X., Wang, L., 2012{\natexlab{a}}. Probabilistic interconnection
  between interdependent networks promotes cooperation in the public goods
  game. J. Stat. Mech. 2012, P11017.

\bibitem[{Wang et~al.(2012{\natexlab{b}})Wang, Szolnoki, and
  Perc}]{wang_z_epl12}
Wang, Z., Szolnoki, A., Perc, M., 2012{\natexlab{b}}. Evolution of public
  cooperation on interdependent networks: The impact of biased utility
  functions. EPL 97, 48001.

\bibitem[{Wang et~al.(2012{\natexlab{c}})Wang, Szolnoki, and
  Perc}]{wang_z_srep12}
Wang, Z., Szolnoki, A., Perc, M., 2012{\natexlab{c}}. If players are sparse
  social dilemmas are too: Importance of percolation for evolution of
  cooperation. Sci. Rep. 2, 369.

\bibitem[{Wang et~al.(2012{\natexlab{d}})Wang, Szolnoki, and
  Perc}]{wang_z_pre12b}
Wang, Z., Szolnoki, A., Perc, M., 2012{\natexlab{d}}. Percolation threshold
  determines the optimal population density for public cooperation. Phys. Rev.
  E 85, 037101.

\bibitem[{Wang et~al.(2013{\natexlab{a}})Wang, Szolnoki, and
  Perc}]{wang_z_srep13}
Wang, Z., Szolnoki, A., Perc, M., 2013{\natexlab{a}}. Interdependent network
  reciprocity in evolutionary games. Sci. Rep. 3, 1183.

\bibitem[{Wang et~al.(2013{\natexlab{b}})Wang, Szolnoki, and
  Perc}]{wang_z_srep13b}
Wang, Z., Szolnoki, A., Perc, M., 2013{\natexlab{b}}. Optimal interdependence
  between networks for the evolution of cooperation. Sci. Rep. 3, 2470.

\bibitem[{Yamagishi et~al.(2012)Yamagishi, Horita, Mifune, Hashimoto, Li,
  Shinada, Miura, Inukai, Takagishi, and Simunovic}]{yamagishi_pnas12}
Yamagishi, T., Horita, Y., Mifune, N., Hashimoto, H., Li, Y., Shinada, M.,
  Miura, A., Inukai, K., Takagishi, H., Simunovic, D., 2012. Rejection of
  unfair offers in the ultimatum game is no evidence of strong reciprocity.
  Proc. Natl. Acad. Sci. USA 109, 20364--20368.

\end{thebibliography}

\end{document}